\documentclass{ws-rmp}
\usepackage{graphicx}
\usepackage{enumerate}
\usepackage{amsmath}
\usepackage{amssymb}
\usepackage{amsfonts}
\usepackage{color}
\usepackage{pstricks}
\usepackage{color}
\newtheorem{thm}{Theorem}[section]

\newtheorem{defin}[thm]{Definition}
\begin{document}

\markboth{R.  Gait\'an, M.G. Morales}
{The Feynman integral path a Henstock integral: a survey and open problems}

%
\catchline{}{}{}{}{}
%

\title{The Feynman integral path a Henstock integral: a survey and open problems}

\author{R. Gait\'an\footnote{
Departamento de F\'isica, FES-Cuautitl\'an, UNAM,  Estado de M\'exico 54770, M\'exico.}}

\address{Departamento de F\'isica, FES-Cuautitl\'an, UNAM,\\  Estado de M\'exico 54770, M\'exico,\footnote{Departamento de F\'isica,  Facultad de Estudios Superiores-Cuautitl\'an, Universidad Nacional Aut\'onoma de  M\'exico.}\\ 
\email{rgaitan@unam.mx\footnote{rgaitan@unam.mx.}} }

\author{M. G. Morales}

\address{Department of Mathematics and Statistics, Faculty of Science,
Masaryk University,\\ Kotl\'a\v{r}sk\'a 2,  Brno, 611 37, Czech Republic.\\ maciasm@math.muni.cz  }

\maketitle

\begin{history}
\received{(Day Month Year)}
\revised{(Day Month Year)}
\end{history}

\begin{abstract}

The  Feynman path  integral is defined over the space  $\mathbb{R}^T$ of all possible paths; it has been a powerful tool to develop Quantum Mechanics. The  absolute value of Feynman's integrand is not integrable, then Lebesgue integration theory could not be used by Feynman. However, it exists formally as a Henstock integral (which does not require the measure concept) and is a suitable alternative to the ordinary integrals that normally appear in path integrals. Feynman proved the equivalence of his theory with the traditional formulation of Quantum Mechanics, since his path integral satisfies Schr\"odinger’s equation. On the other hand, Feynman's  path integral is related to the diagrams of Feynman. For the application of this integral in Feynman's diagrams it is necessary to exchange the integral $\int_{\mathbb{R}^T}$ and the series. 
We discuss the impossibility to exchange  the integral and the sum, considering  integral of Henstock and the version of Dominated Convergence Theorem. Even it has not been proved through the several mathematical formalisms that have been used.
\end{abstract}



\section{Introduction}
\label{sec1}

The Schr\"odinger equation describes the evolution of a state function for a particle of a constant mass $m$ moving in Euclidean space $\mathbb{R}^{d}$ in the presence of a potential
$V (x)$. The state $\psi$ at time $t = 0$ gives the initial condition for the equation and allows one to uniquely determine the state $\psi$ function at all subsequent times:

\begin{equation*}
\frac{\partial\psi}{\partial t}=i\left[\frac{1}{2m}\Delta-V\right]\psi
\end{equation*}
\begin{equation*}
\psi|_{t=0}=\varphi,
\end{equation*}
where,  $\hbar=1$, $\Delta$ is the Laplacian operator in $\mathbb{R}^{d}$ and $V:\mathbb{R}^{d}\rightarrow\mathbb{R}$ is a measurable
function.
Suppose that, we measure successive positions of a particle in one-dimensional space separated by a small time-interval $\epsilon$, denote them by $x_1, x_2, x_3 . . $. Letting the intervals
between measurements $\epsilon$ get smaller and smaller, we would expect the sequence $x_1,x_2, x_3 . . .$  converges to a path of the particle, represented by a function of time $x(t)$.

By Perturbation Theory is obtained that the state function exists as a limit of product of operators, see \cite{Gelfand}, \cite{Kato} and \cite{Nelson}, among others. Thus, according to Feynman's intuition and Perturbation Theory we have  
\begin{eqnarray}\psi(x,t)&=&\lim_{\epsilon\rightarrow 0}\int_{R}e^{i\sum S(x_{i+1},x_{i})}...\frac{dx_{i}}{A}\frac{dx_{i+1}}{A}...\nonumber \\
&=&\lim_{n\rightarrow\infty}\left(\frac{mn}{2\pi i t}\right)^{nd/2}\underbrace{\int_{\mathbb{R}^{d}}...\int_{\mathbb{R}^{d}}}_{n\  times}e^{i\sum [\frac{m}{2}\frac{(x_j-x_{j-1})^{2}}{(t/n)^{2}}-V(x_{j})]^{t/n}}\varphi(x_{n})dx_1...dx_{n}\nonumber\\
&=&\lim_{n\rightarrow\infty}\left( K_{m}^{t/n}M_{V}^{t/n}\right)^{n}\varphi(x)\label{incremento}\end{eqnarray}
where $\frac{1}{A}$ is the normalization factor. However, there exist some difficulties; for example: 

\begin{arabiclist}
 \item the normalization  ``constant'' has a meaning for every finite ``$n$'', but it
becomes infinite as $n$ approaches infinity;
 \item it is
well known that a path of a Brownian particle is continuous but with probability one nowhere differentiable function and $S(x_0,x_{1},...,x_n;t)$ is a classical
action. It means that $x(t)$ must be differentiable;
 \item and  finally, $\lim_{n\rightarrow\infty}\prod_{j=1}^{n}dx_{j}$ corresponds to some measure on a space of all
possible paths or histories, but this product is infinite, thus defined this way,
the measure  does not have firm mathematical meaning.
\end{arabiclist}
Using the Wiener measure and defining the solution to the
Schr\"odinger equation as a limit of a sequence of functional integrals, it was shown that the corresponding measure in the case of the
Feynman path integral does not exist because it fails to be positive and to have the
property of countable additivity, see for example \cite{Se}. 

These kind of difficulties are saved if  the Feynman path integral is understood as a Henstock integral over  all paths, $\mathbb{R}^{T}$, and this integral justifies the  Feynman's intuition, to interpret
 the state function as ``a sum of complex contributions, one from each path in the region'', see \cite{Feynman1}. Furthermore, Henstock's integral is given by Riemann sums, is not absolutely integral and is not needed to introduce a new measure, see \cite{Muldowney1}, \cite{Mu}, \cite{Muldowney2}. Thus, Henstock integral
and non-absolute integrability of the Fresnel integrals provide  a mathematically rigorous definition
supporting Feynman’s intuitive derivations and a representation of the
Feynman’s path integral as a functional. Muldowney gives only local in spacetime
solutions.  In \cite {Nathanson}  is given a global  physical solution to the non-relativistic Schr\"odinger equation in the form of a unitary one-parameter group in $L^2(\mathbb{R}^n)$.
 

One motivation of this work is to show that there is a mathematical foundation such that Feynman's  path integral is supported, \cite {Nathanson}, \cite{Ca}, \cite{Pmuldowney}, \cite{Muldowney1}, \cite{Mu}, \cite{Muldowney2} and \cite{Mi}.
Moreover, this integral has important implications in different branches of science  see for example, \cite{bonotto}, \cite{Muldowney3} and \cite{Muldowney4}, among others. Despite the fact that Feynman's  path integral is well defined according to  the Henstock integral, it has not been formally justified, for example, the application in diagrams of Feynman. We will provide some mathematical tools to discuss this issue. 

 
In the next section we will give the basic definitions and notation of the integral of Henstock in finite dimension.

\section{Definitions: finite-dimensional case}
\label{sec2}
We follow the notation from \cite{Se} and \cite{Mu}  in order to present
basic definitions of the Henstock integral theory; we  introduce the definitions of the integral in the finite-dimensional case, $\mathbb{R}^{n}$ with $n\geq 1$. 

\begin{defin}\label{cell} A cell $I$ in $\mathbb{R}^{n}$ consists in the product $I=I(N)=I_{1}\times...\times I_{n}$, where $N=\{1,...,n\}$ and each $I_{j}$ can have the form \[(-\infty,a),\ \ \ [u,v],\ \ \ (b,\infty],\ \ \ \text{ or}\ \ \ (-\infty,\infty).\]
\end{defin}
The collection of all cells in $\mathbb{R}^{n}$ is denoted by $I(\mathbb{R}^{n})=\{I(N)\}$. Let $\overline{\mathbb{R}}=\mathbb{R}\cup \{-\infty. \infty\}$.
The definition of the Henstock integral is given by Riemman sums, thereby the cells must be related  or associated to points $x\in \mathbb{\overline{ R}}^{n}$, $x=(x_{1},...,x_{n})$.

\begin{defin}Let    $I$ be a cell in $\mathbb{R}^{n}$. The cell is associated to $x\in\mathbb{\overline{  R}}^{n}$ if for each $j=1,...,n$

\begin{arabiclist}
\item $x_{j}=-\infty$, if $I_{j}=(-\infty,a]$
\item $x_{j}=u$ or $x_{j}=v$, if $I_{j}=[u,v]$
\item $x_{j}=\infty$, if $I_{j}=(b,\infty)$
\item $x_{j}=-\infty$ or $x_{j}=\infty$, if $I_{j}=(-\infty,\infty)$
\end{arabiclist}
\end{defin} 

The association condition means that the point $x$ should be in the interior or on the boundary of $I$.

\begin{defin}A \textbf{gauge} in $\mathbb{R}^{n}$ is a positive function $\delta$ defined for $x\in\mathbb{\overline{ R}}^{n}$. An attached point-cell pair $(x,I)$ of $\mathbb{R}^{n}$ is $\delta-fine$ if, for each $j$, the pair $(x_{j},I_{j})$ is $\delta-fine$ in $\mathbb{R}$; that is, 
\[a<\frac{-1}{\delta(x)};\ \ \ u-v<\delta(x);\ \ \ \text{or}\ \ \ b>\frac{1}{\delta(x)},\ \ \text{respectively.}\]
A \textbf{partition} of $\mathbb{R}^{n}$ is a finite collection $\mathcal{P}$ of disjoint cells whose union is $\mathbb{R}^{n}$. A \textbf{division} $\mathcal{D}$ of $\mathbb{R}^{n}$ is a finite collection  of associated point-cell pairs $(x,I)$ whose cells  form a partition of $\mathbb{R}^{n}$. Given a gauge $\delta:\mathbb{\overline{ R}}^{n}\rightarrow\mathbb{R}^{+}$, a division $\mathcal{D}$ is $\delta-fine$ if each $(x,I)\in\mathcal{D}$ is $\delta-fine$, it is denoted by $\mathcal{D}_{\delta}$. 
\end{defin}
In general, an integrand in $\mathbb{R}^{n}$ is a point-cell function $h(x,I)$ defined in the product $\mathbb{\overline{ R}}^{n}\times I(\mathbb{R}^{n})$ to $\mathbb{C}$ (or $\mathbb{R}$), in particular it can be a product $f(x)g(I)$. 

\begin{defin} A function $h(x,I)$ is integrable in $\mathbb{R}^{n}$ with integral $\alpha=\int_{\mathbb{R}^{n}}h(x,I)$ if, given  $\epsilon>0$ there exists a gauge function $\delta=\delta(\epsilon)$ in $\mathbb{\overline{ R}}^{n}$ such that, for each $\delta-fine$ division $\mathcal{D}_{\delta}$ of $\mathbb{R}^{n}$, the corresponding Riemann sums satisfies
$$\left|\alpha-\mathcal{D}_{\delta}\sum h(x,I)\right|<\epsilon,$$ 
where $\mathcal{D}_{\delta}\sum h(x,I)= \sum_{(x,I)\in \mathcal{D}_{\delta}}h(x,I)$. \end{defin}
We follow \cite{Se} and \cite{Mu}  to extend this definitions over the space of all possible paths.
%
%
\section{Definitions: infinite-dimensional case.}
\label{sec3}
Now, we introduce the basic definitions to the infinite-dimensional case. Let $T$ be an interval in $(0,\infty)$. The notation $\mathbb{R}^{T}$ defines  the set of real-valued functions defined on $T$, $(x_{t})_{t\in T}$; also can be understood as $x$ is an element of the Cartesian product $$\prod \{\mathbb{R},t\in T\}=\mathbb{R}^{T}$$

According to the finite-dimensional case, first let us define cells in $\mathbb{R}^{T}$. $\mathcal{N}=\mathcal{N}(T)$ denotes the class of all finite subsets $N$ of $T$. 

\begin{defin} Let  $N$ be a finite set in $\mathcal{N}$, such that $t_1<t_2<...<t_n$. A cell in $\mathbb{R}^{T}$ is
$$I[N]=I(N)\times \mathbb{R}^{T\backslash N}.$$

\end{defin}
According to the interpretation of $\mathbb{R}^{T}$, now $\mathbb{R}^{T\backslash N}$ is the set of all real-valued functions defined on $T\backslash N$. For the integral is not necessary that $t_1<t_2<...<t_n$; however, it is important to define random variation, see \cite{Mu}.
It is helpful to emphasize the ``restricted'' dimensions $N$ of the cylindrical interval $I\subset \mathbb{R}^{T}$, written as $I[N]$. While with round brackets, is a
finite-dimensional interval, $I(N)\subset \mathbb{R}^{n}$. The collection of all cells in $\mathbb{R}^{T}$ is denoted by $I(\mathbb{R}^{T})=\{I[N]:N\in\mathcal{N}\}$.

\begin{defin}
A \textbf{partition} of $\mathbb{R}^{T}$ is a finite collection $\mathcal{P}$ of disjoint cells $I[N]$ whose union
is $\mathbb{R}^{T}$.
\end{defin}
 
In finite dimension the pairs $((x_{1},x_{2},...,x_n),I_1,\times I_2\times I_n)$ on which an integrand $h$ operates are called associated. This concept will be used to define associated point-cell in $\mathbb{R}^{T}$.

\begin{defin} It is said that $(x,N,I[N])$ is associated in $\mathbb{R}^{T}$ if the corresponding finite-dimensional pair $(x(N),I(N))$ is associated in the finite-dimensional space $\mathbb{R}^{n}$.
\end{defin}

Thus, an integrand in $\mathbb{R}^{T}$ may be expressed as $h(x,N,I[N])$, where the triple $(x,N,I[N])$ is associated if the pair $(x(N),I(N))$ is  associated in $\mathbb{R}^{n}$. Note that $x(N)=(x(t_{1}),...,x(t_n))$, $x_{i}=x(t_i)$ and $I_{j}=I_{t_{j}}$ for each $i=1,..,n$.

In addition,  a further condition is imposed  on the lengths of the restricted edges $I_j$: the dimension sets $N$ (or sets of restricted dimensions) of partitioning cells $I[N]$ should include some minimal set of dimensions $L(x)$, for each associated $x$. That is, we require $ L(x)\subset N$, where $L(x)$ can be made successively larger, just as $\delta(x)$ is made successively smaller in forming Riemann sums. Then, a gauge in $\mathbb{R}^{T}$ is considered as a pair of mappings $(\delta,L)$. For more details see \cite{Se}, \cite{Mu}, \cite{Muldowney2}, \cite{RH1}, \cite{RH2}, \cite{RH4}, \cite{RH5}.

\begin{defin} A gauge $\gamma$ in $\mathbb{R}^{T}$ is a pair of mappings $(\delta, L)$ such that
\begin{arabiclist}
\item $L:\overline{\mathbb{R}}^{T}\rightarrow\mathcal{N}(T)$, it means, $x\rightarrow L(x)\in \mathcal{N}=\mathcal{N}(T)$
\item $\delta:\overline{\mathbb{R}}^{T}\times \mathcal{N}(T) \rightarrow (0,\infty)$, it means, $(x,N)\rightarrow \delta(x,N)$.
\end{arabiclist}
\end{defin}

\begin{defin} Let $N=\{t_1,...,t_n\}\in \mathcal{N}(T)$. An associated triple $(x,N,I[N])$ is $\gamma-fine$ if, $L(x)\subseteq N$ and $(x_{i}, I_{i})$ is $\delta-fine$ for $t_{j}\in N$, $1\leq j\leq n$.
\end{defin}

\begin{defin}
A \textbf{division} of  $\mathbb{R}^{T}$ is a finite collection $\mathcal{D}$ of point-cell pairs
$(x,I[N])$ such that the corresponding $(x,N,I[N])$ are associated, and the cells $I[N]$ form a partition $\mathcal{P}$ of $\mathbb{R}^{T}$. 
\end{defin}

\begin{defin} Let  $\gamma = (L, \delta)$ be a gauge function. A division $\mathcal{D}$ is a $\gamma-fine$ if
each $(x,I[N])\in \mathcal{D}$ is $\gamma-fine$. In that case, we can denote the $\gamma-fine$ division $\mathcal{D}$
by $\mathcal{D}_{\gamma}$.
\end{defin}

Suppose that $h = (x, N, I )$ is a real- or complex-valued function
of associated elements $(x, N, I[N])$ in the infinite-dimensional domain $\mathbb{R}^{T} $ and $\mathcal{D} = \{(x, I[N])\}$ is a division of $\mathbb{R}^{T}$, then the corresponding Riemann sum for $h$ is
$$(\mathcal{D})\sum h:=\sum_{(x,I[N])\in\mathcal{D}}h(x,N,I[N])=\sum\{ h(x,N,I[N]):(x,N,I[N])\in\mathcal{D}\}.$$

\begin{defin}\label{defi} A function $h$ of associated triples $(x,N,I[N])$ is integrable on $\mathbb{R}^{T}$, with integral
$$\alpha=\int_{\mathbb{R}^{T}}h(x,N,I[N]),$$
if, given $\epsilon>0$ there exists a gauge $\gamma$ in $\mathbb{R}^{T}$ so that, for each $\gamma-fine$ division $\mathcal{D}_{\gamma}$ of
$\mathbb{R}^{T}$, the corresponding Riemann sum satisfies
$$|\alpha-(\mathcal{D}_{\gamma})\sum h|<\epsilon.$$
\end{defin}

An integrand in $\mathbb{R}^{T}$ might then take the form
$h(x,N,I[N]) = f(x)|I[N]|$ for some point function $f$, where $|I[N]|$ is given by 

\begin{equation}|I[N]|= \left \{ \begin{matrix} \prod_{j=1}^{n} (v_{j}-u_{j}) & \mbox{if }& I_{j}=(u_{j}-v_{j}] \\
 0 & \mbox{Otherwise }&\mbox{}\end{matrix}\right..\label{volumen1}
 \end{equation}
Note that Definition \ref{defi} makes sense only if there exits at least one $\gamma-fine$ division on $\mathbb{R}^{T}$, which is valid for finite and infinite-dimensional case, see \cite{Mu}  and \cite{RH3}. Since $\mathbb{R}^{T}$ is unbounded in each dimension it is
easy to see that constant functions $f(x)$ are not integrable on $\mathbb{R}^T$ with respect
to $|I[N]|$ unless, for instance, $f(x)$ is equal to zero for all $x$. That is why this approach will be based on an application of the Henstock integration
technique to the Fresnel integrands.
%
%
\section{Fresnel Integral: finite-dimensional density and probability distribution functions}
\label{sec4}
First, we will introduce \textit{one-dimensional Henstock Fresnel integral} in order to present the infinite-dimensional case.

In \cite{Se}, it was shown how the Feynman's path integral leads to the Wiener kernel, with the purely imaginary diffusion coefficient and how this makes the corresponding measure not countably additive. The main obstacle consists in the exponential with pure imaginary exponents. The improper Riemann integrals of such expressions are called the Fresnel integrals:
\begin{equation}\int_{-\infty}^{\infty}e^{\frac{i}{2}x^{2}}dx.
\label{integral}\end{equation}
It is shown  that the  integral (\ref{integral}) exists as  Henstock integral and is equal to $\sqrt{\frac{2\pi}{-i}}$, see \cite[Theorem 20]{Se}. This result can be generalized to any complex number $c=a+ib$, where $a\leq 0 $, $b\geq 0$ and $c\neq 0$, see \cite{Mu}.\\

Let us  define a complex-valued function $\varphi(x)$ for $x=(x_1,1_2,...,x_n)\in\mathbb{R}^{n}$ as:

$$\varphi(x)=\varphi_{n,\frac{i}{2}}(x)=e^{\frac{i}{2}(x_1^2+x_2^2+...+x_n^2)}.$$

Now we consider a ``volume'' function $\mu$ defined on a set $I(\mathbb{R}^{n})$ of finite-dimensional cells $I(N)=I_1\times...\times I_n$ as
\begin{equation*}\mu(I(N))=|I(N)|= \left \{ \begin{matrix} \prod_{j=1}^{n} (v_{j}-u_{j}) & \mbox{if }& I_{j}=(u_{j}-v_{j}] \\
 0 & \mbox{Otherwise }&\mbox{}\end{matrix}\right..\label{volumen}
 \end{equation*}
 
Consider the function $h(x,I(N))=\varphi(x)\mu(I(N))$. We will refer to its integral as the Henstock Fresnel integral, and
$$\int_{\mathbb{R}^{n}}\varphi(x)\mu(I(N))=\int_{\mathbb{R}^{n}}\prod_{j=1}^{n}e^{\frac{i}{2}x_j^{2}}|I_j|=\left(\sqrt{\frac{2\pi}{-i}}\right)^{n},$$
see \cite{Se}.

We redefine or normalize the Fresnel integrand in order to  the new integral can be considered as a probability distribution function. If a cell $I \in I(\mathbb{R}^n)$ has an associated point $x \in \overline{\mathbb{R}}^{n}\setminus\mathbb{R}^{n}$, that is, $x = (x_1, . . . ,x_n)$ has some
component $x_j = \pm\infty$, then a value zero was assigned to ``volume'' $\mu(I(N)) = |I|$, by
convention. Since  it is needed to construct a probability distribution function nonzero
values are assigned in all cases. Thus, we define the following function:

\begin{equation*}g_n(x)|I|= \left \{ \begin{matrix} 
\left(\sqrt{\frac{-i}{2\pi}}\right)^{n}e^{\frac{i}{2}x_1^2+...+x_n^2}|I_1|...|I_2| & \mbox{if }& x\in\mathbb{R}^{n} \\
\left(\sqrt{\frac{-i}{2\pi}}\right)^{n} \prod_{j=1}^{n}\int_{I_j}e^{\frac{i}{2}x_j^{2}}dx_j  & \mbox{if }& x \in \overline{\mathbb{R}}^{n}\setminus\mathbb{R}^{n} \end{matrix}\right..\label{densidad Rn}
 \end{equation*}
For example, if $n=1$ and $J=(u,\infty)$, $u>0$, then 
\begin{eqnarray}
g_1(x)|J|&=&\sqrt{\frac{-i}{2\pi}}\left[\frac{1}{2}\int_{-\infty}^{\infty}e^{\frac{i}{2}y^{2}}dy-\int_{0}^{u}e^{\frac{i}{2}y^{2}}dy\right]\nonumber\\
&=&\frac{1}{2}-\sqrt{\frac{-i}{2\pi}}g_1(x)|(0,u)|.\nonumber
\end{eqnarray} 
 The function $g_n(x)|I|$ is integrable in $\mathbb{R}^{n}$ in Henstock sense and $$\int_{\mathbb{R}^{n}}g_n(x)|I|=1.$$ For more details see \cite{Mu}, \cite{Se}. If we consider $g_n(x)|I|$ as an $n$-dimensional Henstock Fresnel density function, then we define its associated $n$-dimensional probability distribution function on cells, 
 \begin{equation*}G_n(I)= \left \{ \begin{matrix} 
\left(\sqrt{\frac{-i}{2\pi}}\right)^{n}\int_I e^{\frac{i}{2}x_1^2+...+x_n^2}|I| & \mbox{if }& x\in\mathbb{R}^{n} \\
\left(\sqrt{\frac{-i}{2\pi}}\right)^{n} \prod_{j=1}^{n}\int_{I_j}e^{\frac{i}{2}x_j^{2}}dx_j  & \mbox{if }& x \in \overline{\mathbb{R}}^{n}\setminus\mathbb{R}^{n} \end{matrix}\right..\label{distribucion Rn}
 \end{equation*}
 
A {\bf figure} in $\mathbb{R}^{n}$ is the union of a finite number of cells, denoted by $E$. The collection of all
figures in $\mathbb{R}^{n}$ is denoted as $E(\mathbb{R}^{n})$. 

In fact, $G_n$ is defined over the collection of figures $E(\mathbb{R}^{n})$, and this is finitely additive on disjoint figures and $\int_{\mathbb{R}^n} G_n(I)=1$, it means, this is a probability distribution function, see \cite[Theorem 154]{Mu}.\\
 
In the classical sense, this function would not be considered as a probability distribution function, because it can take negative or complex values. However, in the formulation of Quantum Mechanics, complex-valued functions play the
role of probability distribution functions. Thus, in this context, we can refer to $G_n(I)$
as a probability distribution function. 
%
%
\section{Fresnel Integral:  infinite-dimensional density and probability distribution functions}
\label{sec5}

By similar considerations over $\mathbb{R}^{T}$ the Fresnel
density and probability distribution functions are defined for each $I[N]\in I(\mathbb{R}^{T})$.

Suppose that $(x(N),I[N])$ are associated in $\mathbb{R}^{T}$, with $N=\{t_1,...,t_n\}\in\mathcal{N}$, $I_j=I_{t_{j}}$, and $x_{j}=x(t_j)\in\overline{\mathbb{R}}$ for each $t_{j}\in N$ and each $N\in \mathcal{N}$. Then the finite-dimensional object $(x(N),I(N))$ corresponds to the infinite-dimensional object $(x_{T},I[N])$ and 

 \begin{equation*}g^{T}(x(N))|I[N]|= \left \{ \begin{matrix} 
\left(\sqrt{\frac{-i}{2\pi}}\right)^{n} e^{\frac{i}{2}x_1^2+...+x_n^2}|I[N]| & \mbox{if }& x\in\mathbb{R}^{T} \\
\left(\sqrt{\frac{-i}{2\pi}}\right)^{n} \prod_{j=1}^{n}\int_{I_j}e^{\frac{i}{2}x_j^{2}}dx_j  & \mbox{if }& x \in \overline{\mathbb{R}}^{T}\setminus\mathbb{R}^{T} \end{matrix}\right..\label{densidad RT}
 \end{equation*}
\begin{equation*}
G^{T}(I[N])=\left(\sqrt{\frac{-i}{2\pi}}\right)^{n}\prod_{j=1}^{n}\int_{I_j}e^{\frac{i}{2}x_{j}^{2}}dx_{j},
\end{equation*}
where $g^{T}:\overline{\mathbb{R}}^{T}\times \mathcal{N}\rightarrow\mathbb{C}$ and $G^{T}:E(\mathbb{R}^{T})\rightarrow\mathbb{C}$, and  $E(\mathbb{R}^{T})$ denotes the collection of all
figures in $\mathbb{R}^{T}$ (figures in $\mathbb{R}^{T}$ are similarly defined). In fact, $G^{T}$ is a probability distribution function, see \cite[Theorem 154]{Mu}. Moreover, any distribution function is integrable, see \cite[Theorem 9]{Mu}.

However, the theory for the Feynman's  path integral  requires to consider the Fresnel
integrals in the incremental form, see the state function (\ref{incremento}). The increments in the variables are just translations
that do not change significantly any results stated before. Thus from now on, we will
use the following definitions for the density and probability distribution functions, $g^{T}(x(N))|I[N]|$ and $G^{T}([N])$, respectively
\begin{equation}g^{T}(x(N))|I[N]|= \left \{ \begin{matrix} 
\left[\prod_{j=1}^{n}\left(\sqrt{\frac{-i}{2\pi t_{j}-t_{j-1}}}\right) e^{\frac{i}{2}\frac{(x_j-x_{j-1})^{2}}{(t_{j}-t_{j-1})}}\right]|I[N]| & \mbox{if }& x\in\mathbb{R}^{T} \\
\prod_{j=1}^{n}\left(\sqrt{\frac{-i}{2\pi t_{j}-t_{j-1}}}\right) \int_{I_j}e^{\frac{i}{2}\frac{(x_j-x_{j-1})^{2}}{(t_{j}-t_{j-1})}}dx_{j}  & \mbox{if }& x \in \overline{\mathbb{R}}^{T}\setminus\mathbb{R}^{T} \end{matrix}\right..\label{densidad-incrementos RT}
 \end{equation}
\begin{equation}
G^{T}(I[N])=\prod_{j=1}^{n}\left(\sqrt{\frac{-i}{2\pi t_{j}-t_{j-1}}}\right) \int_{I_j}e^{\frac{i}{2}\frac{(x_j-x_{j-1})^{2}}{(t_{j}-t_{j-1})}}dx_{j},
\label{distribution RT}\end{equation}
where $T\subset \mathbb{R}^{+}$ and $N=\{t_{1},t_{2},...,t_n\}\in\mathcal{N}(T)$ with $0=t_0<t_1<t_2<,...,t_n$. Here the differences $x_j - x_{j-1} = x(t_j)- x(t_{j-1})$ are the increments or transitions of the incremental or transitional Fresnel integrand.

By \cite[Theorem 168]{Mu} the function $G^{T}$ in the expression (\ref{distribution RT}) is integrable on $\mathbb{R}^{T}$ and defines a distribution function.
 
\begin{thm}\label{teo}
Let be $f(x,N)$ a real- or complex-valued function. If either $f(x,N)G^{T}(x[N])$ or $f(x,N)g^{T}(x(N))|I|$ is integrable on  $\mathbb{R}^{T}$, then the other is also integrable and the integrals are equal. The functions $G^{T}(x[N])$ and $g^{T}(x(N)|I(N)|$  are defined in (\ref{densidad-incrementos RT}) and (\ref{distribution RT}), respectively.
\end{thm} 
 
\begin{example} Let be $g^{*}(x(N))$ the complex conjugate of $g^{T}(x(N))$ (expression (\ref{densidad-incrementos RT})). By Theorem \ref{teo}, the function $g^{*}(x(N))g^{T}(x(N))|I[N])|$ is not integrable since $$g^{*}(x(N))g^{T}(x(N))=|g^{T}(x(N))|^{2}$$is always positive, so no cancellation takes place when Riemann sums are formed.
Then it is easy to see that the Riemann sums diverge. Thus, $g^{*}(x(N))$ is not $G^{T}$ integrable.  
\end{example} 
 
Another example of an integrable  function but non absolute integrable is the following.
\begin{example}\label{ejemplo} Let us consider the function $g_{0}(x,N)$ as, $$g_{0}(x,N)=e\left(\frac{i}{2}\sum_{j=1}^{n}\frac{(x_{j}-x_{j-1})^{2}}{t_{j}-t_{j-1}}\right)\prod_{j=1}^{n}(2\pi i (t_{j}-t_{j-1}))^{-1/2},$$ when $x\in \mathbb{R}^{T}$ 
\end{example}
Note that $g_{0}(x,N)=g^{T}(x(N))$ if $x\in\mathbb{R}^{T}$. We redefine $g^{T}$ as $g_{0}$ because we emphasize the free particle case, it means $V=0$. 
In  \cite[Proposition 68]{Muldowney1} it is shown that $g_0$ is \textit{Henstock Fresnel} integrable. To prove that $g_0$ is not absolutely integrable  it is enough to observe that the function $e^{iu^{2}}$ is not Lebesgue integrable, according to the proof of Proposition 68 from \cite{Muldowney1}. Let us write $$e^{iu^{2}}=\cos(u^{2})+i\sin(u^{2}).$$
Since $$\int_{-\infty}^{\infty}e^{iy^{2}}dy=\sqrt{i\pi},$$
we have $$\int_{[0,\infty)}\cos(u^{2})du=\int_{[0,\infty)}\sin(u^{2})du=\frac{1}{2}\sqrt{\frac{\pi}{2}}.$$
These integrals exist as extended Riemann integrals, but not as Lebesgue integrals. The graphs
of $cos (u^2)$, $sin(u^2)$ are oscillate periodically
with constant amplitude 2 but with period decreasing to zero as $u\rightarrow\infty$.

\section{Problem statement}

 In \cite{Mu} and \cite{Muldowney2} it is shown that, under certain conditions, the state function of  Schr\"odinger equation exists as ``a sum of complex contributions, one from each path in the region''. According to the Feynman intuition, in the sense of Henstock integral, it means, if $V$ is continuous and $T$ is a bounded interval in $(0,\infty)$, then the state function $\psi$ is given as
\begin{eqnarray} \psi(\xi, \tau)&=&\int_{\mathbb{R}^{T}}e\left(-i\sum_{j=1}^{n}V(x_{j-1})(t_{j}-t_{j-1})\right)\nonumber\\
&&e\left(\frac{i}{2}\sum_{j=1}^{n}\frac{(x_{j}-x_{j-1})^{2}}{t_{j}-t_{j-1}}\right)\prod_{j=1}^{n}(2\pi i (t_{j}-t_{j-1}))^{-1/2}|I[N]|,
\end{eqnarray} where the displacement $\xi$ at
time $\tau$. In other words,

\begin{equation*}\psi(\xi,\tau)=\int_{\mathbb{R}^{T}}\left(e^{-i\int_{T}V(x(t),t)dt}e^{\frac{i}{2}\int_{T}(\frac{dx}{dt})^{2}dt}\right)\prod_{t\in T}\frac{\delta x(x)}{\sqrt{2\pi i dt}},\label{1}
\end{equation*} where $\prod_{t\in T}\delta x(x)$ is the volume of the cell  $|I|$. Moreover, in \cite{Muldowney1} it was proved that  if $V=0$, then 

\begin{eqnarray*}\psi_{0}(\xi,\tau)&=&\int_{\mathbb{R}^{T}}g_{0}(x,N)|I[N]|\nonumber\\&=&\int_{\mathbb{R}^{T}}e\left(\frac{i}{2}\sum_{j=1}^{n}\frac{(x_{j}-x_{j-1})^{2}}{t_{j}-t_{j-1}}\right)\prod_{j=1}^{n}(2\pi i (t_{j}-t_{j-1}))^{-1/2}|I[N]|\nonumber\\&=& \left(2\pi i(\tau-\tau')\right)^{-1/2}e\left(\frac{i/2(\xi-\xi')^{2}}{\tau-\tau'}\right).\end{eqnarray*}
On the other hand, if $e^{-i\int_{T}V(x(t),t)}dt$ is expressed as a series $$\sum_{r=0}^{\infty}(r!)^{-1}\left(-i\int_{T}V(x(t),t)dt\right)^{r},$$
then we have 
\begin{equation}
\psi(\xi,\tau)=\int_{\mathbb{R}^{T}}\left(\sum_{r=0}^{\infty}(r!)^{-1}\left(-i\int_{T}V(x(t),t)dt\right)^{r} e^{\frac{i}{2}\int_{T}(\frac{dx}{dt})^{2}dt}\right)\prod_{t\in T}\frac{\delta x(x)}{\sqrt{2\pi i dt}}.
\label{serie}
\end{equation}
Assuming that we can interchange the order of the integral and the series in the expression (\ref{serie}) (this is assumed in the Perturbation Theory, see \cite{Feynman}), then $\psi=\sum_{r=0}^{\infty}\psi_{r}$, where each $\psi_{r}$ is given by the recursive sequence: 
\begin{equation*}
\psi_{r}=\frac{i}{r}\int_{T}v(x(s_r),s_r)\psi_{r-1}ds_r,
\end{equation*}
where $$\psi_{0}=\int_{\mathbb{R}^{T}}e^{i\int_{T}\frac{1}{2}\frac{dx^{2}}{dt}}\prod_{t\in T}\frac{\delta x(t)}{\sqrt{2\pi i dt}}.$$
Each term $\psi_r$ of the series has a visual representation as a Feynman's diagram and corresponds to a particular physical phenomenon, where there is interaction among the particles. For example, $$\psi_{1}=-i\int_{\mathbb{R}^{T}}\int_{T}V(x(s),s)ds e^{i\int_{T}\frac{1}{2}\frac{dx^{2}}{dt}}\prod_{t\in T}\frac{\delta x(t)}{\sqrt{2\pi i dt}},$$ is understood as the sum over all possible paths of free particle amplitude. However, each path is weighted by $V(x(s),s)$, that is, before and after time $ s $, the paths are of a free particle, then there is a perturbation in time $s$.\\

\subsection{Limit under the sign of the integral}
Recently the Henstock integral over infinite-dimensional spaces  has been developed. Thus there are generalized versions of Monotone Convergence Theorem and  Dominated Convergence Theorem, see \cite{Mu} and \cite{Muldowney2}. On the other hand, we would like to  justify formally the exchange of integral and series in (\ref{serie}). However, we try to prove that with the mathematical tool that gives us the integral of Henstock there are difficulties and it is not possible to make this justification. Conditions  to guarantee the limit under the sign of the integral are presented.
\begin{defin}\label{convegencia} Let  $(h_{m})$ be a sequence of real- or complex-valued functions. $h_m$ converges boundedly to $h$ if there exists a positive-valued function $\beta(x,N,I[N])$,  such that, given $\epsilon>0$, for each associated   $(x,N,I[N])$ there is a gauge $\gamma_{0}$ and an integer $m_{0}=m_{0}(x,N,I[N]) $ such that $$|h_{m}(x,N,I[N])-h(x,N,I[N])|<\epsilon \beta(x,N,I[N]),$$
for every  $m\geq m_{0}$ and for every  $\gamma_{0}-fine$ $(x,N,I[N])$.
\end{defin} 

That is, the ``pointwise convergence'', $|h_{m}(x,N,I[N])-h(x,N,I[N])|<\epsilon$ for $m>m_{0}$ and every $(x,N,I[N])$, is not sufficient to guarantee a type result Dominated Convergence Theorem on the space of the Henstock functions integrable in $\mathbb{R}^{T}$, see \cite{Muldowney2}. Thus, we present a result to guarantee the limit under the integral sign, in the context of bounded convergence over $\mathbb{R}^{T}$, see \cite{Mu}.

\begin{thm}
Suppose the sequence $h_{j}(x,N,I[N])$ converges to  $h(x,N,I[N])$ in $\mathbb{R}^{T}$ in the sense of the  Definition \ref{convegencia}. Then  $h$ is integrable if and only if there exist a ball $B_{1}$ of arbitrarily small radius and correspondingly,  a $\gamma$ and integers $p=p(x,N,I[N])$, depending on $B_{1}$ such that 
$$D_{\gamma}\sum h_{m(x,N,I[N])}(x,N,I[N])\in B_{1}$$ for all choices of $m\leq p $ and for all $\gamma-fine$ $D_{\gamma}$. 
\end{thm}It means that the limit under the sign of the integral holds.
Thus, we define the following sequence of functions, for $m=0,1,2,...$

\begin{eqnarray}h_{m}(x,N,I[N])&=&\sum_{r=0}^{m}(r!)^{-1}\left(-i\sum_{j=1}^{n}V(x_{j-1})(t_{j}-t_{j-1})\right)^{r}\nonumber\\&& \cdot e\left(\frac{i}{2}\sum_{j=1}^{n}\frac{(x_{j}-x_{j-1})^{2}}{t_{j}-t_{j-1}}\right)\cdot\prod_{j=1}^{n}(2\pi i (t_{j}-t_{j-1}))^{-1/2}|I[N]|,\nonumber\end{eqnarray}
where $|I[N]|$ is defined by the expression (\ref{volumen1}).\\
Suppose that $$\psi(\xi,\tau) = \sum_{r=0}^{\infty} \left(\int_{\mathbb{R}^{T}}(r!)^{-1}\left(-i\int_{T}V(x(t),t)dt\right)^{r} e^{\frac{i}{2}\int_{T}(\frac{dx}{dt})^{2}dt}\right)\prod_{t\in T}\frac{\delta x(x)}{\sqrt{2\pi i dt}},$$ that is, the exchange of the integral and the series holds. Thus, according to the previous theorem, $h_{m}$ converges boundedly to $$h(x,N,I[N])=e\left(-i\sum_{j=1}^{n}V(x_{j-1})(t_{j}-t_{j-1})\right)e\left(\frac{i}{2}\sum_{j=1}^{n}\frac{(x_{j}-x_{j-1})^{2}}{t_{j}-t_{j-1}}\right)$$ $$\prod_{j=1}^{n}(2\pi i (t_{j}-t_{j-1}))^{-1/2}|I[N]|.$$
Therefore, there is a positive and integrable function $\beta(x,N,I[N])$ so that, given $\epsilon >0$, for each associated  $(x,N,I[N])$ there is a gauge $\gamma_{0}$ and an integer  $m_{0}$ such that $$|h_{m}(x,N,I[N])-h(x,N,I[N])|<\epsilon\beta
(x,N,I[N]),$$ for every $m\geq m_{0}$ and for every  $\gamma_{0}-fine$ $(x,N,I[N])$.\\

There are other versions of the Dominated Convergence Theorem in the sense of the Henstock integral see \cite{Mu} and \cite{Muldowney2}. However, it was decided to use this version since it is not required an increasing succession, thus avoiding the problem of comparing to the elements $ h_m $, since these are functions in addition to complex values.

Continuing with the above assumption, we have:\\
$|h_{m}(x,N,I[N])-h(x,N,I[N])|=$
$$\left|\sum_{r=m+1}^{\infty}(r!)^{-1}\left(-i\sum_{j=1}^{n}V(x_{j-1})(t_{j}-t_{j-1})\right)^{r}e\left(\frac{i}{2}\sum_{j=1}^{n}\frac{(x_{j}-x_{j-1})^{2}}{t_{j}-t_{j-1}}\right)\prod_{j=1}^{n}(2\pi i (t_{j}-t_{j-1}))^{-1/2} |I[N]|\right|$$ \hspace{5cm} $<\epsilon\beta(x,N,I[N]).$\\
On the other hand,\\
$|h_{m}(x,N,I[N])-h(x,N,I[N])|=$
$$\left|\sum_{r=m+1}^{\infty}(r!)^{-1}\left(-i\sum_{j=1}^{n}V(x_{j-1})(t_{j}-t_{j-1})\right)^{r}e\left(\frac{i}{2}\sum_{j=1}^{n}\frac{(x_{j}-x_{j-1})^{2}}{t_{j}-t_{j-1}}\right)\prod_{j=1}^{n}(2\pi i (t_{j}-t_{j-1}))^{-1/2}|I[N]|\right|$$
\hspace{5cm} $<\epsilon |g_{0}(x,N,I[N])|,$ \\
for $m$ large enough. Here 
$$g_{0}(x,N,I[N])=e\left(\frac{i}{2}\sum_{j=1}^{n}\frac{(x_{j}-x_{j-1})^{2}}{t_{j}-t_{j-1}}\right)\prod_{j=1}^{n}(2\pi i (t_{j}-t_{j-1}))^{-1/2}|I[N]|, $$ in addition, it is proved that this function is not an absolutely integrable function, see Example \ref{ejemplo}.\\

\subsection{Partial ordering: Cone} The mathematical technique to introduce partial order in a Banch space is the concept of cone, see \cite{Dajun}.

\begin{defin} Let $E$ be a Banach space and  $P$ be a nonempty closed convex set. $P$ is called cone if: 
\begin{arabiclist}
\item If $x\in P$ y $\lambda\geq 0$, then $\lambda x\in P$
\item If $x$ and $-x$ belong to $P$, then $x=0$, where 0
 denotes the zero element in $E$.
 \end{arabiclist}
\end{defin}

In this case 0 denotes the function incidentally zero. 

\begin{defin} Every cone $P$ in $E$ defines an order relation $\leq$ in  $E$ as follows:
\[x\leq y \text{\ \ if\ \ } y-x\in P.\]
\end{defin}

We must consider a functions set $E$ such that $|g_{0}|$, $\beta$  belong to  $E$. Moreover, $E$ must be a Banach space, it means that there exists a norm over $E$; since $|g_{0}|$ is not integrable this norm must not be induced by the integral.

The usual way is to define $P=\{g(x,N,I):\ g(x,N,I[N])\geq 0,\ \ for\ \ each \ \ \gamma_{0}-fine \ \ (x,N,I[N])\}$ with some norm (it is easy to show that $P$ is a nonempty convex set, but we require  the norm in order to know if $P$ is a closed set). \\

Suppose that $\beta- |g_{0}|\in P$ for each $\gamma_{0}-fine$ $(x,N,I[N])$, which implies that $|g_{0}|\leq \beta$, for every  $\gamma_{0}-fine$ $(x,N,I[N])$, nevertheless
$|g_{0}|$ is not integrable, which leads us to a contradiction since for the convergence on integrable functions in $\mathbb{R}^{T}$ it's necessary that
 $\beta$ be integrable. This implies that it is  not possible exchange the integral $\int_{\mathbb{R}^{T}}$ and the series. Note that this exchange is necessary to apply Feynman's Diagram, see (\ref{serie}) and \cite[Perturbation Theory]{Feynman}. In order to justify the exchange, under Henstock integral (since Feynman's path integral exists as Henstock integral), we must assume the bounded convergence. Thus, with this tool is not possible exchange the integral $\int_{\mathbb{R}^{T}}$ and the series.  \\
 
  In the case that, $|g_{0}|-\beta\in P$ for each $\gamma_{0}-fine$ $(x,N,I[N])$ implies that $\beta \leq |g_0|$, for every  $\gamma_{0}-fine$ $(x,N,I[N])$, here we need to introduce the supremum in Banach spaces in order to give a contradiction. This case is being analyzed with supremum concept.


Suppose that $P$ is a cone in a Banach space $E$ and $P$ defines a partial order in $E$. Let $D\subset E$ be nonempty. An element $z\in E$ is called a supremum of $D$ if it satisfies the two conditions:
\begin{arabiclist}
\item $x\leq z$ for all $x\in D$;
\item $x\leq y$ for all $x\in D$, then $z\leq y$.
\end{arabiclist}
\vspace{.3cm} 
  

On the other hand, it is well known that the set of Henstock-Kurzweil integrable functions, $HK([a,b])$ ($[a,b]\subset \overline{\mathbb{R}}$), is not complete respect to  Alexiewicz norm, $||f||_{A}=\sup_{x}\left|\int_{a}^{x}f(t)dt\right|$, see \cite{Sw}. Thus it is natural to study and consider its completion respect to Alexiewicz norm, see \cite{Bongiorno} and \cite{Ta}. Moreover, there exists a norm on $HK([a,b])$ under which it is a Banach space, see \cite[Proposition 9]{Gutierrez} and this norm is not natural, that is, is not induced by the integral, see \cite{Honig}. Thus,  we could assume that the space of integrable functions in $\mathbb{R}^{T}$ with a norm induced by the integral, is not a Banach space.\\

We consider  important to study properties of the function space $$H=\left\{h:\mathbb{R}^{T}\times \mathcal{N}(T)\times I(\mathbb{R}^{T})\rightarrow\mathbb{C}: \left|\int_{\mathbb{R}^{T}}h\right|<\infty \right\}.$$ Note that  $N\in \mathcal{N}(I)$ fixes the dimension of the cell $I(N)$ and therefore of $I[N]$. In the case that $T$ contains a finite number of elements   $T=\{t_{1},...,t_{n}\}$, then for $1\leq j\leq n$ and arbitrary cells $I_j$ in $\mathbb{R}^{\{t_{j}\}}=\mathbb{R}$, denote an arbitrary cell $I\in I(\mathbb{R}^{T})$ of $\mathbb{R}^{T}=\mathbb{R}^{\{t_{1},...,t_{n}\}}$ by $$I_1\times,...,\times I_{n}.$$
When $n=1$, $\mathbb{R}^{T}=\mathbb{R}$  the cells take the form from Definition \ref{cell}.

Let us fix $n=1$ and consider the set of step functions defined in  $\mathbb{R}$, denoted by $S$. Note that $$S\subset H,$$ because  $s:\mathbb{R}\rightarrow\mathbb{R}$ is a step function, if exists a compact subinterval os $\mathbb{R}$ such that $s$ is a step function on $J$ and $s(x)=0$ if $x\in \mathbb{R}-J$. In fact, if $n=1$ and $f$ is Lebesgue integrable, then $f$ is integrable in Henstock sense; in  \cite{Mu} is called Riemann-complete integrable. Thus, $\textrm{ Card}( H)\geq c$, where $c$ is the cardinality of real numbers and therefore, $\textrm{Dim}(H)\geq c$. \\

We denote $H'=\{F:E(\mathbb{R}^{T})\rightarrow\mathbb{C}: F\ \ is \ \ additive\ \ function, |f(x,N,I[N])-F(I[N])|\ \ is \ \ integrable\ \ and\ \ \int_{\mathbb{R}^{T}}|f-F|<\epsilon\}$. Then $F$ is the indefinite integral of $f$, $F(E)=\int_{E}f$.
It is easy to show that $\textrm{ Card}( H)= \textrm{Card}(H')$, see \cite[Theorem 19]{Mu}.
 In the case of $F(\mathbb{R}^{T})=1$, then $F$ is  a distribution function. Therefore, the space of distribution functions is contained in $H'$. \\


We would like to show that $\textrm{Card}(H)\leq c,$ then $\textrm{Dim}(H)=c$. By \cite[Corollary 2.4]{Kruse} it is provided of  a Banach space structure to $H$. We do not know if this norm is induced by the integral. Nevertheless,
 in the case $\mathbb{R}^{n}$, it is possible to show that there exists no natural Banach space norm on Henstock integrable functions over $\mathbb{R}^{n}$ to $\mathbb{R}$, see \cite[Theorem 2.7]{Honig}, which makes us think that this result can be extended over $H$, then the norm that provided $H$ of the Banach space structure would have the possibility of not being induced by the integral. If it is possible to provide $H$ with a Banach space structure, then we can use the cones and partial ordering. Thus, to be able to get at a contradiction with respect to $| g_0 |$ must be the supremum of $(h_{m})$, so $|g_{0}|\leq \beta$ and therefore, there is not a justification about the  exchange of the integral and the series, in the sense of the Henstock integral. 
\section{Conclusions} Although the Henstock integral gives a mathematical formalism to the  Feynman integral path which is an important tool in applications in Quantum Mechanics, to our knowledge, the mathematical instruments  seem to be not enough to justify particular applications  as Feynman's diagrams.
\section*{Acknowledgments}
This work is partially supported by CONACYT-SNI. M.G.M. and R.G.L acknowledge support by  project PAPIIT-IN113916 and program DGAPA-UNAM of  postdoctoral fellowship.


\begin{thebibliography}{9}









\bibitem{Gelfand}I. M. Gel'fand and A. M. Jaglom, Integration in functional spaces and its applications in quantum physics, {\it J. Mathematical Phys.} {\bf 1} (1960)  48--69.
\bibitem{Kato} T. Kato, Fundamental properties of Hamiltonian operators of Schr\"odinger type, {\it Trans. Amer. Math. Soc.} {\bf 70} (1951)   195--211.
\bibitem{Nelson} E. Nelson, Feynman integrals and the Schr\"odinger equation, {\it J. Mathematical Phys.} {\bf 5} (1964)  332--343.
\bibitem{Se} E. S. Nathanson, Path integration with non-positive distributions and applications to the Schrodinger equation, Ph. D. Thesis, The University of Iowa. ProQuest LLC, Ann Arbor, MI, (2014).
\bibitem{Feynman1}R. P. Feynman, Space-time approach to non-relativistic quantum mechanics, {\it Rev. Modern Physics} {\bf 20} (1948) 367--387.
\bibitem{Muldowney1} P. Muldowney, {\it A general theory of integration in function spaces, including Wiener and Feynman integration}  (John Wiley \& Sons, Inc., New York, 1987).
\bibitem{Mu} P. Muldowney, {\it A modern theory of random variation. With applications in stochastic calculus, financial mathematics, and Feynman integration} (John Wiley \& Sons, Inc., Hoboken, NJ, 2012). 

\bibitem{Muldowney2} P. Muldowney, Feynman's path integrals
and Henstock's non-absolute integration, {\it J. Appl. Anal.} {\bf 6} (2000) 1--24.

\bibitem{Nathanson} E. S. Nathanson and  P. E. T. Jorgensen, A global solution to the Schr\"odinger equation: from Henstock to Feynman. {\it J. Math. Phys.} {\bf 56} (2015),15 pp. 

\bibitem{Ca} S. S. Cao, The Henstock integral for Banach-valued functions, {\it Southeast Asian Bull. Math.} {\bf 16} (1992) 35--40.
\bibitem{Pmuldowney} R. Henstock, P. Muldowney and  V. A. Skvortsov, Partitioning infinite-dimensional spaces for generalized Riemann integration, {\it Bull. London Math. Soc.} {\bf 38} (2006) 795--803.

\bibitem{Mi} M. Pardy, The Generalized Feynman Integral, {\it Internat. J. Theoret. Phys.} {\bf 8} (1973)  31--37.

\bibitem{bonotto} E. M. Bonotto, M. Federson and P. Muldowney, A Feynman-Kac
solution to a random impulsive equation of Schr\"odinger type, {\it Real Anal. Exchange} {\bf 36}  (2010/11)  107--148.
\bibitem{Muldowney3} P. Muldowney, The Henstock and the
Black-Scholes theory of derivative asset pricing, {\it Real Anal. Exchange} {\bf 26}  (2000/01) 117--131.

\bibitem{Muldowney4} P. Muldowney, Financial valuation and
the Henstock integral, {\it Semin\'ario Brasileiro de An\'alise,} Sao Paulo {\bf 60} (2000b) 79--108.


\bibitem{RH1} R. Henstock, {\it Linear analysis} (Butterworths, London, 1968).

\bibitem{RH2}  R. Henstock, Integration in product spaces, including Wiener and Feynman integration,
{\it Proc. London Math. Soc.} {\bf 27} (1973) 317–344.

\bibitem{RH4} R. Henstock, {\it The general theory of integration} (Oxford University Press, Oxford, 1991).
\bibitem{RH5}  R. Henstock, The construction of path integrals, {\it Mathematica Japonica} {\bf 39} (1994) 15–18.



\bibitem{RH3} R. Henstock, {\it Lectures on the theory of integration} (World Scientific, Singapore, 1988).

\bibitem{Dajun} G. Dajun, J. C. Yeol  and Z. Jiang, {\it Partial Ordering Methods in no Linear Problems}  (Nova Science Publishers, Inc., Hauppauge, NY, 2004).


\bibitem{Feynman}R. P. Feynman and A. R. Hibbs, {\it Quantum Mechanics and Path Integrals} (Dover Publications, Inc., Mineola, NY, 2010).



\bibitem{Sw} C. Swartz, {\it Introduction to gauge integrals} (World Scientific Publishing Co., Singapore, 2001).

\bibitem{Bongiorno} B. Bongiorno and T. V. Panchapagesan, On the Alexiewicz
topology of the Denjoy space, {\it Real Anal. Exchange} {\bf 21}   (1995/96) 604--614.
\bibitem{Ta} E. Talvila,  The distributional Denjoy integral, {\it Real Anal. Exchange} {\bf 33} (2008) 51--82.

\bibitem{Gutierrez} L. Guti\'errez M\'endez, J. Escamilla Reyna,
M. C\'ardenas, and J. Estrada Garc\'ia, The Closed Graph Theorem and the Space of Henstock-Kurzweil
Integrable Functions with the Alexiewicz Norm,  {\it Abstr. Appl. Anal.} {\bf 2013} Art. ID 476287, 4 pp.


\bibitem{Honig} C. S. H\"onig, There is no natural Banach space norm on the space of Kurzweil-Henstock-Denjoy-Perron integrable functions in {\it 30 Sem. Bras. An\'alise}, Sao Paulo  (1989), pp. 387-397.




















\bibitem{Kruse} A. H. Kruse, Badly incomplete normed linear spaces, {\it Math. Z.} {\bf 83} (1964)  314--320.












\end{thebibliography}
\end{document}